\documentclass[12pt]{book}

\usepackage[dvips]{graphicx,color}
\usepackage{latexsym,amsmath}
\usepackage{makeidx,tsukuba}
\newcommand{\Ibar}{{\mathcal I} \kern -0.5em\raise 0.25ex \hbox{--}}
\newif\ifpdf
\ifx\pdfoutput\undefined
\pdffalse 
\else
\pdfoutput=1 
\pdftrue
\fi

\makeauthorindex
\makeindex

\begin{document}

\BookTitle{\itshape The 28th International Cosmic Ray Conference}
\CopyRight{\copyright 2003 by Universal Academy Press, Inc.}
\pagenumbering{arabic}

\chapter{
Gravitational Radiation -- Observing the Dark and Dense Universe}

\author{%
%
%
B.S. Sathyaprakash\\
{\it School of Physics and Astronomy, Cardiff University, Cardiff, CF24 3YB, UK}\\
}

\section*{Abstract}
Astronomical observations in the electromagnetic window -- microwave, radio and optical --
have revealed that most of the Universe is dark. The only reason
we know that dark matter exists is because of its gravitational influence on 
luminous matter. It is plausible that a small fraction of that dark matter is clumped, and 
strongly gravitating.  Such systems are potential sources of gravitational radiation that can
be observed with a world-wide network of gravitational wave antennas. Electromagnetic astronomy
has also revealed objects and phenomena -- supernovae, neutron stars, black holes and the
big bang --  that are without doubt extremely strong emitters of the radiation targeted by the
gravitational wave interferometric and resonant bar  detectors. In this talk I will highlight why 
gravitational waves arise in Einstein's theory, how they interact with
matter, what the chief astronomical sources of the radiation are, and in which 
way by observing them we can gain a better understanding of the dark and dense Universe. 

\section{Introduction}
Einstein's theory of gravity admits wave-like solutions that are in many ways 
similar to electromagnetic radiation but with important differences, two among 
them being most crucial: Universal, but weak, interaction and non-linearity. The former
property has profound consequences: It implies that one cannot infer the influence of
gravitational waves by watching an isolated particle in space, one would need at least
two well-separated particles, just as one would in Einstein's gedanken lift-experiment 
to infer the presence of the Earth's gravitational field. The weakness of the gravitational
interaction means that on the one hand it will be very difficult to observe them, but
on the other the radiation carries the true signature of the emitting source, be it
the core of a neutron star or a supernova, the quasi-normal mode oscillations of
a black hole, or the birth of the Universe, thereby making it possible to 
observe phenomena and objects that are not directly accessible to the 
electromagnetic, neutrino or the cosmic-ray window.  The latter property, namely
the non-linearity of the waves, means that the waves interact with the source resulting
in a rich structure in the shape of the emitted signals. Therefore, gravitational 
radiation should also facilitate both quantitatively and 
qualitatively new tests of Einstein's theory including the measurement of the 
speed of gravitational waves, and hence (an upper limit on) the mass of the 
graviton, polarisation states of the radiation, non-linear effects of general
relativity untested in solar system or Hulse-Taylor binary pulsar observations,
uniqueness of axisymmetric spacetimes, and so on.

The influence of gravitational radiation on an antenna can be characterized 
by a dimensionless amplitude which is a measure of the deformation caused
by the wave as it passes through the detector. For instance, in an interferometric
antenna of length $\ell$ a wave of amplitude $h$ causes a change in length
$\delta \ell = h \ell/2.$ Typical astronomical events, say a binary black hole
merger at 100 Mpc, would have an amplitude $h\sim 10^{-23}$ at a frequency  
$\sim 100$~Hz and such events can be expected to occur at a once every few years.
Nearer and/or stronger events could produce amplitudes that are
several orders of magnitudes larger, but their event rate would be too low.
The technology needed to observe such tiny amplitudes
has become available only in the past decade or so.  Many resonant bars and 
interferometers are currently taking data near their design sensitivity in
the range $10^{-21}$--$10^{-23}$ and should soon be in a position to observe 
some of the most violent phenomena in the Universe.

In this talk I will begin with a brief overview of gravitational wave (GW) theory
and the interaction of the waves with matter and how that is used in the construction 
of the detectors.  The main focus of this talk will be the astronomical sources,
tests of general relativity, and astrophysical and cosmological measurements 
afforded by GW observations.  Sec.~\ref{sec:conventions} lists our choice 
of units and the conventions used in making estimates of source strengths.

\def\vecx{\mathbf{x}}
\def\vecr{\mathbf{r}}
\def\gab{g_{\alpha\beta}}
\def\Gab{G_{\alpha\beta}}
\def\Tab{T_{\alpha\beta}}
\def\hab{h_{\alpha\beta}}
\def\hbarab{\overline{h}_{\alpha\beta}}
\def\eab{\eta_{\alpha\beta}}
\def\Eab{\eta^{\alpha\beta}}
\def\gsim{\mathrel{ \rlap{\raise 0.511ex \hbox{$>$}}{\lower 0.511ex \hbox{$\sim$}}}}
\def\lsim{\mathrel{ \rlap{\raise 0.511ex \hbox{$<$}}{\lower 0.511ex \hbox{$\sim$}}}}

\section{Gravitational wave theory - A brief overview}
Newtonian gravity is described by a scalar potential $\varphi(t,\, \vecx),$
which obeys the Poisson equation, $\nabla^2\varphi(t,\, \vecx) = 4\pi \rho(t,\, \vecx),$
where $\rho(t,\, \vecx)$ is the density distribution, whose 
formal solution is given by
\begin{equation}
\varphi(t,\, \vecx) = \int \frac{\rho(t,\, \vecx')\, {\rm d}^3 x' }{|\vecx - \vecx'|}.
\end{equation}
The key point is that because the potential satisfies a Poisson equation there
are no retardation effects. Since the same time $t$ appears
both on the LHS and the RHS in the above equation, any change in the
distribution of density at the source point $\vecx'$ would {\it instantaneously}
change the potential at the remote field point $\vecx.$ 

\subsection{Wave equation}
In Einstein's theory the metric components of the background spacetime are
the gravitational potentials and they satisfy a ``wave'' equation and hence there will
be retardation effects. This is explicitly seen in the linearized version of
Einstein's equations. Under the assumption of weak gravitational fields one
can assume that the background metric $\gab$ of spacetime to be only 
slightly different from the Minkowski metric $\eab = {\rm Diag}(-1,\, 1,\, 1,\, 1):$
$\gab = \eab + \hab.$ Here $\hab$ is a part of the metric that describes the departure of 
the spacetime from flatness. For weak gravitational fields, 
there exists a coordinate system in which 
each component of $\hab$ is numerically small compared to unity, i.e.\ $|\hab| \ll 1.$ 
For this reason $\hab$ is termed as the {\it metric
perturbation.} Furthermore, even when the source that produces
the weak field is relativistic, Einstein's 
equations $\Gab = 8 \pi \Tab,$ where $\Gab$ is the Einstein
tensor and $\Tab$ is the energy-momentum tensor, reduce, on keeping only
terms linear in $\hab,$ to a set of wave equations for the metric perturbation:
$\Box \hbarab = 16 \pi \Tab,$ where $2 \hbarab \equiv 2 \hab - \eab h^\mu_\mu,$ and
$\Box$ is the wave operator: $\Box \equiv \Eab \partial_\alpha \partial_\beta.$
These equations have the formal solution
\begin{equation}
\hbarab(t,\vecx) = 4 \int \frac{\Tab(t-|\vecx - \vecx'|,\, \vecx')\, {\rm d}^3 x'}{| \vecx - \vecx'|}.
\end{equation}
In this solution, the metric perturbation at the field point $\vecx$ at the time $t$ is
determined by the configuration of the source $\Tab$ at a retarded time $t-|\vecx-\vecx'|/c$
($c$ being the speed of light),
and hence disturbances in the source travel only at a finite speed. Indeed, any
non-stationary source $\Tab$ will give rise to wave-like solutions for the potentials
$\hbarab,$ which extract energy, momentum and angular-momentum from the source, 
propagating at the speed of light and have other attributes similar to electromagnetic waves.

\subsection{Polarisation states and principle of detection}
Although to begin with there are 10 independent components of the metric, because
the theory is covariant under general coordinate transformations and invariant under
gauge transformations, one can make a choice of coordinate system and gauge such 
that only two {\it independent} components of the metric are non-zero.
Therefore, just as in electromagnetic theory, there are only two independent polarisations
of the field, denoted as $h_+$ (h-plus) and $h_\times$ (h-cross). 

\begin{figure}
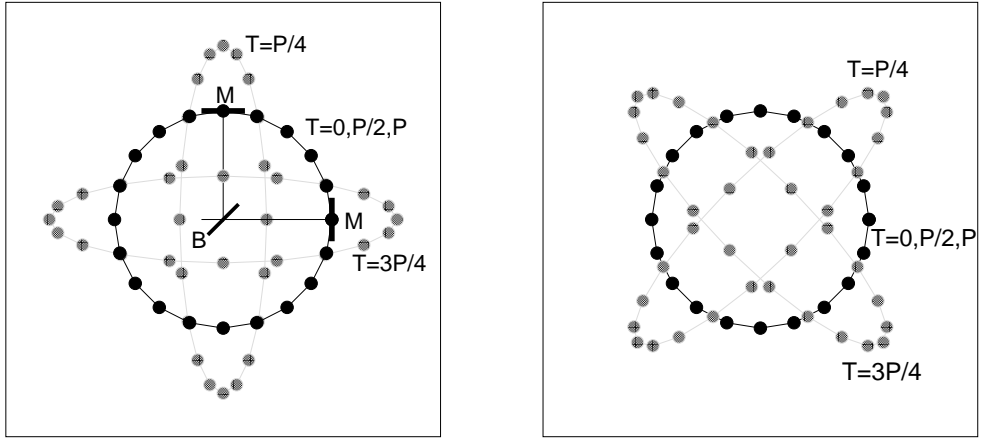

  \begin{center}
    \includegraphics[height=2.3in]{Plus.eps}
\hskip 1.3 cm
    \includegraphics[height=2.3in]{Cross.eps}
  \end{center}
  \vspace{-0.5pc}
\caption{The response of a circular ring of {\it free} beads to waves
of plus- (left) and cross-polarisation (right) as a monochromatic signal of 
period $P.$ The ring (radius $C$) is continuously 
deformed into an ellipse (semi-major axis $(1+h/2)C,$ semi-minor axis $(1-h/2)C,$)
after one quarter of the period.  The configuration of the beads
after a time $T = P/4, P/2, 3P/4$ is shown.  Also depicted are the 
locations of the beam-splitter and mirrors that are freely 
suspended inside a vacuum tube in an interferometric detector.}
\label{fig:ring}
\end{figure}

When a wave of plus- or cross-polarisation is incident 
perpendicular to a plane containing a circular ring of 
beads, the ring is deformed in the manner shown in Fig.\ \ref{fig:ring} 
Monitoring the distance from the centre of the ring to the beads at the ends of 
two orthogonal radial directions can best measure the deformation of the ring. 
This is the principle behind a laser interferometer antenna wherein highly 
reflective mirrors (losses $\sim 10^{-5}$) are freely suspended (quality factors 
$\sim 10^6$)  at the ends of two orthogonal arms (length $\ell \sim \rm km$) 
inside vacuum tanks (pressure $\sim 10^{-8}$ mbar) and high power lasers 
(effective power of 10 kW) are used to measure extremely tiny 
strains ($\delta \ell / \ell \sim 10^{-21}$--$10^{-23}$ for transient bursts and 
$10^{-25}$--$10^{-27}$ for continuous wave sources observed over several months),  
in a wide range (10--1000~Hz) of frequency.
Gravitational wave interferometers are quadrupole detectors with 
a good sky coverage. Indeed, an interferometer will have 40\% of the
peak sensitivity over 40\% of the sky area! 
A single antenna, except when it is a spherical resonant detector, 
cannot determine the polarisation state of a {\it transient} wave or the direction
to the source that emits the radiation. 
Interferometers and resonant bars don't measure the two polarisations separately
but rather a linear combination of the two given by:
\begin{equation}
h(t) = F_+ (\theta,\, \varphi,\, \psi) h_+(t) + F_\times(\theta,\, \varphi,\, \psi)  h_\times(t),
\end{equation}
where $F_+$ and $F_\times$ are the antenna patterns.
To infer the direction $(\theta,\, \varphi)$ to the source,
the polarisation amplitudes $(h_+,\, h_\times),$
and the polarisation angle $\psi,$ it is necessary to make five measurements which is
possible with three interferometers: Each interferometer gives a response, say $h_1(t),$ $h_2(t),$
and $h_3(t),$ and one can infer two independent delays, say $t_1-t_2,$ and $t_2-t_3,$
in the arrival times of the transient at the antennas.  Therefore,
a network of antennas, geographically widely separated so as to 
maximize the time delays and hence improve directionality,
is needed for GW observations.  Moreover, detecting the same 
event in two or more instruments helps to remove the non-Gaussian and non-stationary 
backgrounds, while adding a greater degree of confidence to the detection of an event.
In the case of continuous waves the motion of the detector relative to the source 
causes a Doppler modulation of the response which can be de-convolved from the 
data to fully reconstruct the wave.

\subsection{Amplitude, luminosity and frequency}
The amplitude $h$ and luminosity ${\cal L}$ of a 
source of GW is given 
in terms of the famous quadrupole formula:
\begin{equation}
h_{mn}(t,\vecr) = \frac{2}{r}\, \ddot {\cal \Ibar}_{mn}(t-r),\ \ \ \ 
{\cal L} = \frac{1}{5} \left< \dddot {\cal \Ibar}_{mn} \dddot {\cal \Ibar}^{mn} \right >,
\end{equation}
where an overdot denotes derivative with respect to time;
angular brackets denote a suitably defined averaging process (say, over a period
of the GW);
${\cal \Ibar}_{mn}$ is the {\it reduced} (or trace-free) quadrupole moment tensor which
is related to the usual quadrupole tensor $I^{mn}\equiv \int T^{00} x^m x^n {\rm d}^3 x,$
via ${\cal \Ibar}_{mn} \equiv I_{mn} - \delta_{mn} I^k_{\ k}/3.$ In simple terms, for a source
of size $R,$ mass $M$ and angular frequency  $\omega,$ located at a distance $r$ from Earth,
\begin{equation}
h \sim \epsilon_h \frac{M}{r}R^2 \omega^2,\ \ \ \ {\cal L} \sim \epsilon_{\cal L} M^2 R^4 \omega^6.
\end{equation}
where $\epsilon_{h,\cal L}$ are dimensionless efficiency factors that depend on
the orientation of the system relative to the observer (in the case of $h$ only) and
how {\it deformed} from spherical symmetry the system is. $\epsilon_{h,\cal L} \sim 1$ for ideally
oriented and highly deformed sources. 
The amplitude of the waves, just as in the case of electro-magnetic radiation, decreases 
as inverse of the distance to the source. However, there is a crucial difference between
EM and GW observations that is worth pointing out: Let $r_l$ be the largest distance from
which an EM or a GW detector can observe standard candles. In the case of electromagnetic telescopes
$r_l$ is limited by the smallest flux observable which falls off as the inverse-square 
of the distance. This is because astronomical EM radiation 
is the superposition of waves emitted by a large number of microscopic sources, each photon
with its own phasing; we cannot follow each wave separately but only a superposition
of many of them. This, of course, is the reason why in conventional 
astronomy the number counts of standard candles increase as $r_l^{3/2}.$ 
In the case of GW, signals we expect to observe are emitted by the coherent bulk motion of large masses
and hence it is possible to observe each cycle of the wave as it passes through the
antenna. Indeed, one can fold many wave cycles together to enhance the
visibility of the signal buried in noise, provided the shape of the signal is known
before hand. Because we can follow the amplitude of a wave the number of 
sources which an antenna can detect increases as $r_l^3.$ 

For a self-gravitating system, say a binary system of two stars of masses $m_1$ and $m_2$
(total mass $M=m_1+m_2$ and symmetric mass ratio $\eta=m_1m_2/M^2$), the linear
velocity $v$ and angular velocity $\omega$ are related to the size $R$ of the system
via Keplar's laws: $\omega^2 = M/R^3,$ $v^2 = M/R.$ It turns out that
the efficiency factors for such a system are
$\epsilon_h = 4\eta\, C,$ $\epsilon_{\cal L} = 32\eta^2/5,$ so that
\begin{equation}
h \simeq 4\eta\,C \frac{M}{r}\frac{M}{R}, \ \ \
{\cal L} \simeq \frac{32}{5} \eta^2 v^{10},
\ \ \ f_{\rm GW} = 2 f_{\rm orb},
\end{equation}
where $C\sim 1$ is a constant that depends on the orientation of the source
relative to the detector, $f_{\rm GW}$ is the GW 
frequency which is equal to twice the orbital frequency 
$f_{\rm orb}$\footnote{For a binary consisting of two equal masses 
the configuration of the system is identical on rotation by 
$\pi,$ rather than $2\pi,$ radians.  This is the reason why 
the frequency of GW is twice the orbital 
frequency.  In general, the wave would contain the orbital frequency 
and its harmonics with twice the orbital frequency being the dominant.}.
The above relations imply that the amplitude of a source is greater 
the more compact it is and the luminosity is higher from a source 
that is more relativistic.  The factor to covert the luminosity 
from $G=c=1$ units to conventional units is ${\cal L}_0 \equiv c^5/G 
\simeq = 3.6 \times 10^{59}$~erg~s$^{-1}$.  Since $v<1,$ ${\cal L}_0$ 
denotes the best luminosity a source could ever have and generally
${\cal L} \ll {\cal L}_0$. 

\subsection{The Hulse-Taylor and other binaries}
\label{sec:Hulse-Taylor-Binary}
As an illustration of these order-of-magnitude estimates let us consider
the Hulse-Taylor binary pulsar PSR 1913+16 \cite{Taylor}. Discovered in 1974 the system consists
of two neutron stars each of mass $1.4 M_\odot,$ in a tight orbit with a 
period $P_b\sim 7.75$~Hrs at a distance of 5~kpc from Earth. The expected 
amplitude, luminosity and frequency of GW are $h \sim 6 \times 10^{-23},$ 
${\cal L} \sim 1.4\times 10^{29}{\, \rm erg \,s^{-1}}$ and
$f_{\rm GW} \sim 7.17 \times 10^{-5}{\rm \ Hz},$ respectively. 
Although the frequency of GW is
beyond the reach of the current ground-based, and future space-based detectors, 
the emission of GW causes the orbital period $P_b=2\pi M/v^3$
to decrease in course of time. Demanding that the energy dissipated into GW should 
be balanced by a loss in the binding energy $E = -\eta M v^2/2,$ i.e.
${\cal L} = -dE/dt,$ we can deduce that 
\begin{equation}
\dot P_b= -\frac {192\pi\eta }{5} \left (\frac{2\pi M}{P_b} \right )^{2/3}.
\end{equation}
In reality, since the binary pulsar is in an eccentricity orbit with
ellipticity $e=0.617,$ the rate of change of the period is larger by a 
factor of 10 giving 
$\dot P_b^{\rm GR}=(-2.4047 \pm 0.00002) \times 10^{-12}{\, \rm s\,s^{-1}}$ 
\cite{WeisbergAndTaylor}.  Thus, general relativity predicts
that the period of the binary should change each year by 
$\Delta P_b \simeq -76$~micro seconds.  By monitoring the binary pulsar 
for over 25 years it has been possible to measure $\dot P_b$ very
accurately \cite{WeisbergAndTaylor}
$\dot P_b^{\rm Obs}=(-2.4086 \pm 0.0052) \times 10^{-12}{\, \rm s\,s^{-1}}$ 
and it agrees with the theoretical estimate to better 
than 0.25\%.  The neutron stars in this system would spiral-in 
towards each other and eventually coalesce in about 300 million years. 
The radiation emitted during the last few minutes before coalescence 
of such systems would be the target of ground-based detectors. 

Since 1974 two more compact binaries that would coalesce within the
Hubble time have been discovered (cf. Table~\ref{table:binaries}), 
B1534+12 \cite{Wolszczan} and J0737-3039 \cite{BurgayEtAl}. 
The binary pulsar J0737-3039 has particularly improved the 
prospect of detecting GW with the upcoming detectors.  \cite{KalogeraEtAl}.
\begin{table}[t]
\caption{The orbital period $P_b,$ eccentricity $e,$ derived masses of the pulsar
$m_p$ and its companion $m_c,$ the measured/expected rate of decay of the
period $\dot P_b$ and time to coalescence $\tau$ of 3 
binary pulsars that would coalesce within the Hubble time.}
\label{table:binaries}
\begin{center}
\begin{tabular}{lrcccr}
\hline
\hline\\[-0.4cm]
Binary    & $P_b$/s & $e$   & $(m_p,\, m_c)$/$M_\odot$ & $\dot P_b$/$10^{-12}$ & $\tau$/Myr\\ 
\hline
B1913+16  & 27907   & 0.617 & $(1.44, 1.39)\cite{Taylor} $ & $-2.40$  & 302  \\
B1534+12  & 36352   & 0.274 & $(1.33, 1.35)\cite{Stairs} $ & $-0.14$ & 2730 \\
J0737-3039& 8835    & 0.0877& $(1.24, 1.35)\cite{BurgayEtAl} $ & $-1.24$ &   86   \\
\hline
\hline\\[-1.0cm]
\end{tabular}
\end{center}
\end{table}

\section{Gravitational Wave Detector Projects}
\begin{figure}[t]
  \begin{center}
    \includegraphics[height=4.5in]{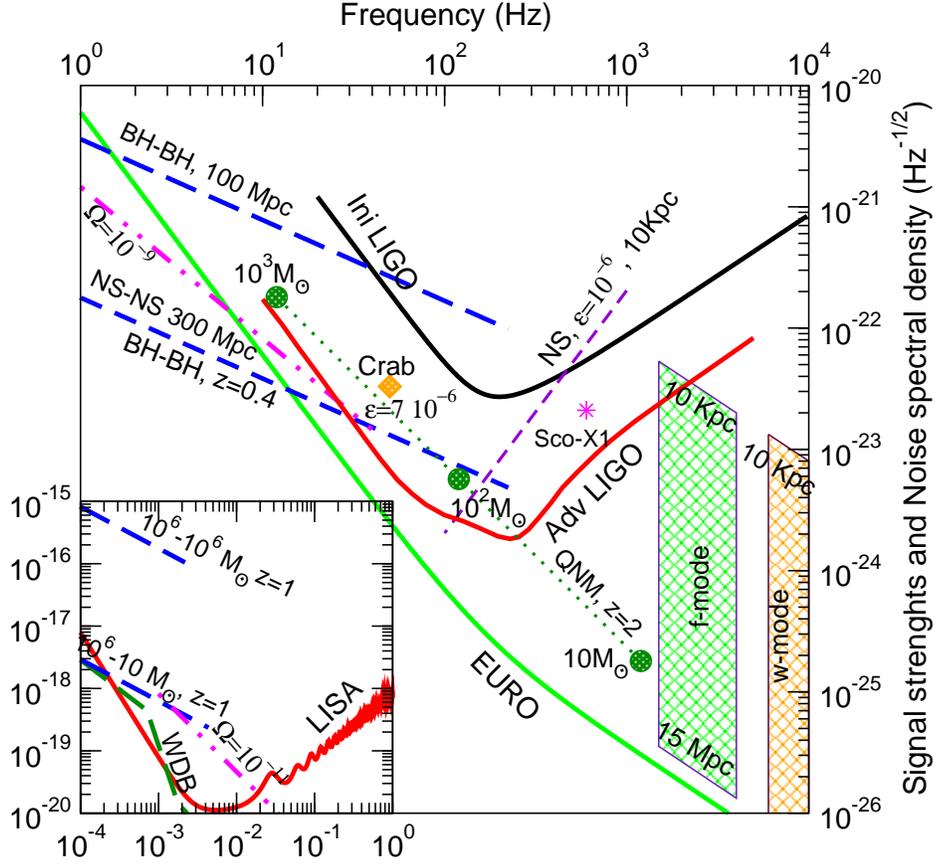}
  \end{center}
  \vspace{-0.5pc}
\caption{Noise PSD (in Hz$^{-1/2}$) of space-based LISA and three
generations of ground-based interferometers, initial and advanced LIGO, and EURO,
respectively.  Also plotted on the same graph are the source strengths 
for archetypal binary, continuous and stochastic radiation in the same 
units.  A source will be visible in a network of 3 interferometers
if it is roughly 5 times above the noise PSD.}
\label{fig:SourcesAndSensitivity}
\end{figure}

There are chiefly two types of GW detectors that are currently
in operation taking sensitive data: (i) {\it resonant bars} and (ii) 
{\it laser interferometers.} The sensitivity of a detector is defined in terms
of the smallest discernible dimensionless strain caused by an astronomical source 
against background noise of the instrument. Because a GW antenna can follow the
phase of GW, the sensitivity of an antenna is given in terms of the amplitude noise
power spectral density as a  function of frequency and is
measured in Hz$^{-1/2}.$ Fig.~\ref{fig:SourcesAndSensitivity} shows in solid 
lines the design sensitivity goals of three generations of ground-based interferometers 
(shown here for the American initial and advanced LIGO, and a possible third 
generation European detector EURO).  The inset shows the same for the space-based LISA.
Also plotted in Fig.~\ref{fig:SourcesAndSensitivity} are source strengths 
to be discussed in Sec.~\ref{sec:sources}
\subsection{Bar detectors}
Resonant bars operate in a narrow band  of 10--50 Hz at a frequency of about 950 Hz 
(see Schutz \cite{SchutzCQG} for a fuller description). 
In a bar detector the vibrations induced in a seismically isolated, cryogenic
Alumninium or Niobium cylindrical bar is amplified using a transducer. There
are currently five such detectors operating around the world, one in Australia (NIOBE),
three in Italy (NAUTILUS \cite{NAUTILUS}, AURIGA \cite{AURIGA}, Explorer\footnote{The Explorer detector is operated
by an Italian group but located in CERN}) and one in the US (ALLEGRO). Bar detectors are
limited by background noises caused by internal thermal noise and the quantum uncertainty
principle.  Current detectors have a strain sensitivity of about $10^{-21}\,\rm Hz^{-1/2}$ 
and are mainly sensitive to supernovae in the neighbourhood of the Milkyway and 
in-band continuous wave sources. 

\subsection{Ground-based Interferometers}
Interferometers operate in a broad band (1 kHz) at a central frequency of 150 Hz 
(see Schutz \cite{SchutzCQG} for a fuller description). 
In a laser interferometric antenna the tidal deformation caused in the two
arms of a Michelson interferometer is sensed as a shift in the fringe pattern at
the output port of the interferometer. The sensitivity of such a detector
is limited at low frequencies (10--40~Hz) by anthropogenic sources and seismic disturbances,
at intermediate frequencies (40--300~Hz) by thermal noise of optical and suspended
components, and at high frequencies ($>300$~Hz) by photon shot noise.  
Three key technologies have made it possible to achieve the current level of sensitivities:
(1)~An optical layout that makes it possible to recycle the laser light exiting
the interferometer and build effective powers that are 1000's of times larger than
the input thereby mitigating the photon shot noise. This technique
allows us to operate the interferometer either in the wide band mode 
(as in Fig.~\ref{fig:SourcesAndSensitivity}), or with 
a higher sensitivity in a narrow band of about 10--50 Hz
centered at a desired frequency, say 300 Hz, but at the cost
of worsened sensitivity over the rest of the band. This
latter mode of operation is called  {\it signal recycling} and
is particularly useful for observing long-lived continuous wave sources.
(2)~Multiple suspension systems
that filter the ground motion and keep the mirrors essentially free from seismic
disturbances. (3)~Monolithic suspensions that help isolate the thermal noise
to a narrow frequency band.  
There are currently six long baseline detectors in operation: The American 
Laser Interferometer Gravitational-Wave Observatory (LIGO) \cite{LIGO}, which is a 
network of three detectors, two with 4 km arms and one with 2 km arms,
at two sites (Hanford, Washington and Livingstone, Louisiana),  the French-Italian
VIRGO detector with 3 km arms at Pisa \cite{VIRGO}, the British-German GEO\,600 \cite{GEO} with
600 m arms at Hannover and the Japanese TAMA with 100 m arms in Tokyo \cite{TAMA}.
Australia has built a 100 m test facility with a plan to
build a km-size detector sometime in the future. 

Plans are well underway both in Europe and the USA to build, by 2008, 
the next generation of interferometers that are 10--15 times more 
sensitive than the initial 
interferometers. With a peak sensitivity of $h \sim 10^{-24}\,{\rm Hz}^{-1/2}$ 
these advanced detectors will be able to detect NS ellipticities as small as $10^{-6}$
in our Galaxy, BH-BH binaries at a redshift of $z\sim 0.5,$ stochastic background
at the level of $\Omega_{\rm GW} \sim 10^{-8}.$ In the longer term, over the next
6 to 10 years, we might see the development of 3rd generation GW antennas.  
The sensitivity of the current and next generation instruments is
still far from the fundamental limitations of a ground-based
detector: The gravity gradient noise at low frequencies 
due to natural (winds, clouds, earth quakes) and anthropogenic causes, and 
the quantum uncertainty principle of mirror position at high frequencies.
A detector subject to only these limitations requires the
development of new optical and cryogenic techniques that form the foundation
of a third generation GW detector in Europe called EURO \cite{EURO},
whose expected noise performance is also shown in Fig.\ref{fig:SourcesAndSensitivity}

\subsection{Space-based Interferometers}
ESA and NASA have resolved 
to place three spacecraft in heliocentric orbit, 60 degrees behind the Earth, 
in an equilateral 
triangular formation of size 5 million km \cite{LISA}. These craft constitute the 
Laser Interferometer Space Antenna (LISA) scheduled to fly in 2011. LISA's 
sensitivity is limited by difficulties with long time-scale ($< 10^{-4}$~Hz)
stability, photon shot-noise ($\sim 10^{-3}$~Hz) and large size ($>10^{-1}$~Hz).
LISA will be able to observe Galactic, extra-Galactic and cosmological point
sources as well as stochastic backgrounds from different astrophysical populations
and perhaps from certain primordial processes. 
In addition to LISA there have been proposals to build an antenna in the
frequency gap of LISA and ground-based detectors. The {\it Deci-Hertz
Interferometer Gravitational-Wave Observatory} (DECIGO) \cite{DECIGO} 
by the Japanese team and the {\it Big-Bang Observer} (BBO), a possible 
follow-up of LISA \cite{BBO}, are aimed as instruments 
to observe the primordial GW background and to answer cosmological 
questions on the expansion rate of the Universe and dark energy.

\section{Sources of gravitational waves}
\label{sec:sources}
Gravitational wave detectors will unveil dark secrets of the Universe
by helping us to study sources in extreme physical environs: strong 
non-linear gravity, relativistic motion, extremely high density, temperature 
and magnetic fields, to list a few.  We shall focus our attention on 
{\it compact objects} (in isolation or in binaries) and {\it stochastic backgrounds}.

\subsection{Compact Binaries}
Compact binaries, consisting of a pair of compact objects (i.e., NS and/or BH),
are an astronomer's standard candles \cite{SchutzNature}: A parameter
called the \emph{chirp mass} ${\cal M} \equiv \eta^{2/3}M,$ completely fixes 
the absolute luminosity of the system. Hence, by observing GW from
a binary we can measure the luminosity distance to the source 
provided the source {\it chirps,} that is
the orbital frequency changes, by as much as $1/T$ during an observational
period $T$.  This feature helps to accurately measure cosmological
parameters and their variation as a function of red-shift.
The dynamics of a compact binary consists
of three phases: (i) {\it inspiral,} (ii) {\it plunge} and (iii) 
{\it merger.}  In the following we will discuss each in turn.

(i) The {\it early inspiral phase:} This is the phase in which the
system spends 100's of millions of years and  the power emitted
in GW is low. This phase can be treated using linearized Einstein's
equations and post-Newtonian theory with the rough energy balance 
between the binding energy and the emitted radiation
(cf. Sec.~\ref{sec:Hulse-Taylor-Binary}). The emitted GW signal 
has a characteristic shape with its amplitude and frequency slowly
increasing with time and is called a {\it chirp} waveform.  Formally, the 
inspiral phase ends at the {\it last stable orbit} (LSO) when the
effective potential of the system undergoes a transition from having 
a well-defined minimum to the one without a minimum, after which stable 
orbits can no longer be supported. This happens roughly
when the two objects are separated by $R\simeq 6\,GM/c^2,$ or when 
the frequency of GW is $f_{\rm LSO} \simeq 4400\, (M_\odot/M)\, \rm Hz.$
The signal power drops as $f^{-7/3}$ and the photon shot-noise in 
an interferometer increases as $f^2$ beyond about 200 Hz so that
it will only be possible to detect a signal in the range from about
10 Hz to 500 Hz (and a narrower bandwidth of 40--300 Hz in initial 
interferometers) during which the source brightens up 
half-a-million fold (recall that the luminosity $\propto v^{10} \propto f^{10/3}$).  
For $M\lsim 10M_\odot,$ inspiral phase is 
the only phase sensed by the interferometers and lasts for a 
duration of $\tau = 5576\, \eta^{-1}\, (M/M_\odot)^{-5/3}\,\rm s,$
starting at 10 Hz. The phase development of the signal is very well 
modelled during this epoch and one can employ matched filtering technique 
to enhance the visibility of the signal by roughly the square root
of the number of signal cycles $N_{\rm cyc} \sim 16 \tau,$ starting
at 10 Hz. Since a large number of cycles are available it is possible
to discriminate different signals and accurately measure the parameters
of the source such as its location (a few degrees each in co-latitude and
azimuth) \cite{JaranowskiEtAl}, mass (fractional accuracy of 0.05--0.3\% in total mass and 
a factor 10 worse for reduced mass, with greater accuracy for NS than BH), 
and spin (to within a few percents) \cite{CutlerAndFlanagan}.  

(ii) The {\it late inspiral,} {\it plunge} and {\it merger phase:} This is
the phase when 
the two stars are orbiting each other at a third of the speed of light and 
experiencing strong gravitational fields with the gravitational
potential being $\varphi = GM/Rc^2 \sim 0.1.$ This
phase warrants the full non-linear structure of Einstein's equations 
as the problem involves strong relativistic gravity, tidal deformation
(in the case of BH-BH or BH-NS) and disruption (in the case of BH-NS and NS-NS)
and has been the focus of numerical relativists \cite{NRBruegmann}
for more than two decades. However, some insights have been gained by
the application of advanced mathematical techniques aimed
at accelerating the convergence properties of post-Newtonian expansions
of the energy and flux required in constructing the phasing of GW \cite{BD1}\cite{BD2}\cite{DIS1}. 
This is also the most interesting phase from the point of view of testing
non-linear gravity as we do not yet fully understand the nature of
the two-body problem in general relativity. Indeed, even the total
amount of energy radiated during this phase is highly uncertain, with
estimates in the range 10\% \cite{FlanaganAndHughes} to 0.07\% \cite{BD2}. 
Since the phase is not well-modelled, it is
necessary to employ sub-optimal techniques, such
as time-frequency analysis, to detect this phase and then use numerical
simulations to gain further insights into the nature of the signal.

(iii) The {\it late merger phase:} This is the phase when the two systems 
have merged to form either a single NS or a BH, settling down to a 
quiescent state by radiating the deformations inherited during the merger. 
The emitted radiation can be computed using perturbation theory and 
gives the quasi-normal modes (QNM) of BH and NS. The QNM carry 
a unique signature that depends only on the mass and spin angular momentum
in the case of BH, but depends also on the equation-of-state (EOS) 
of the material in the case of NS. Consequently, 
it is possible to test conclusively whether or not the newly born object
is a BH or NS: From the inspiral phase it is possible
to estimate the mass and spin of the object quite precisely and use that
to infer the spectrum of normal modes of the BH. The fundamental
QNM of GW from a spinning BH, computed numerically and then fitted, 
is \cite{Echeverria}
$f_{\rm QNM}= 750 [1-0.63 (1-a)^{0.3}](100 M_\odot/M)\,\rm Hz,$ with
a decay time of $\tau = 5.3/[f_{\rm QNM} (1-a)^{0.45}]\,\rm ms,$ where
$a$ is the dimensionless spin parameter of the hole taking values in
the range $[0,\,1].$ The signal will be of the form
$ h(t; \tau, \omega) = h_0 e^{-t/\tau} \cos(\omega t),$ $t\ge 0,$ 
$h_0$ being the amplitude of the signal that depends on how deformed
the hole is.

It has been argued that during the late stages of merger 
the energy emitted in the form of QNM might be as large as 3\% of the 
system's total mass \cite{FlanaganAndHughes}.  By matched filtering,
it should be possible to detect QNM resulting from binary mergers
of mass in the range 60--$10^3\,M_\odot$ at a distance of 
200~Mpc in initial LIGO and from $z\sim 2$ in advanced
LIGO. In Fig.~\ref{fig:SourcesAndSensitivity} filled circles 
(connected by a dotted line) show the
amplitude and frequency of QNM radiation from a source at $z=2,$
and total mass $1000,$ $100$ or $10$ $M_\odot.$ 
Such signals should serve as a probe to test if massive black holes 
found at galactic cores initially formed as small BHs of $10^3\, M_\odot$ 
and then grow by rapid accretion.
Moreover, there is a growing evidence \cite{GerssenEtAl} that globular clusters might 
host BH of mass $M\sim 10^3\,M_\odot$ at
their cores. If this is indeed true then the QNM from activities associated
with such BHs would be observable in the local Universe, depending on
how much energy is released into GW when other objects fall in.
EURO could also observe QNM in stellar mass black holes
of mass $M\sim 10$--$20 M_\odot.$ 

\begin{figure}[t]
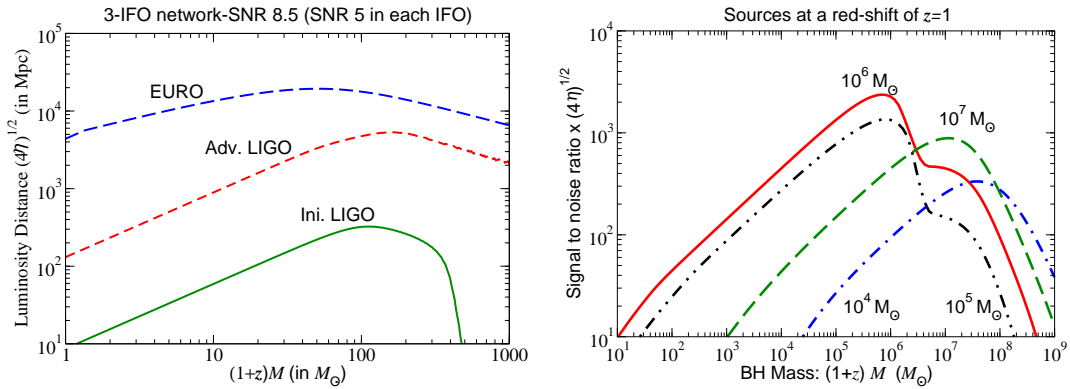

\begin{center}
\includegraphics[height=2.0in]{LIGO-EURO.eps}
\hskip 0.4 true cm
\includegraphics[height=2.0in]{snrvsmassLisa2.eps}
\end{center}
\caption{The span of initial and advanced LIGO and EURO for compact
binary sources when including both the inspiral and merger waveforms
in our search algorithms (left panel).  BH mergers can be seen out to a red-shift
of $z=0.55$ in advanced LIGO and $z=2$ in EURO.  On the right we plot the SNR achieved
by LISA for mergers at $z=1$ of BH of mass as on the $x$-axis with
a MBH of mass is indicated on the plot.}
\label{fig:BinariesSNR}
\end{figure}

The span of an interferometer to binaries varies
with the masses as $\eta^{1/2} M^{5/6},$ greater asymmetry in the masses
reduces the span but larger total mass increases the span. However, for 
$M>100\,M_\odot$ the sensitivity worsens as the seismic
and thermal noise begin to dominate the noise spectrum at lower frequencies.
In Fig.~\ref{fig:BinariesSNR} we have plotted the distance up
to which binaries can be seen as a function of the binary's total mass for
an equal mass system when including both the inspiral and merger part of 
the signal.  This estimate is based on
the effective-one-body approach \cite{BD2}\cite{DIS3} which predicts 
$0.07\%$ of the total mass in the merger waves.  The plunge phase increases 
the signal-to-noise ratio (SNR) by about a factor 1.5 over
the inspiral phase which leads to an increase in the event rate, over that based
purely on the inspiral stage, by a factor of $1.5^3 \simeq 3.4$.
 
\subsubsection{NS-NS binaries}
Double NS can be seen in advanced LIGO to a distance of 300~Mpc 
as shown in Fig.~\ref{fig:SourcesAndSensitivity}
Based on the observed small population of double NS
binaries which merge within the Hubble time (cf. Table\ref{table:binaries}),
Kalogera et al. \cite{KalogeraEtAl} conclude that the 
Galactic coalescence rate is $\sim 1.8 \times 10^{-4}$~yr$^{-1}$ which 
would imply an event rate of NS-NS coalescences of 0.25 and 1500~yr$^{-1},$ in 
initial and advanced LIGO, respectively.  As the spins of NS are
very small ($a\ll 1$) and because the two stars would merge well outside
the LIGO's sensitivity band, the current state-of-the-art theoretical waveforms
\cite{LB2002} 
will serve as good templates for matched filtering. 
However, detailed relativistic hydrodynamical 
simulations (see, e.g. Ref. \cite{KawamuraEtAl}) would be needed to 
interpret the emitted radiation during the coalescence phase, 
wherein the two stars collide to form a bar-like structure 
prior to merger. The bar hangs up over a couple of dynamical time-scales
to get rid of its deformity by emitting strong bursts of GW.
Observing the radiation from this phase should help to deduce the
EOS of NS bulk matter.
Also, an event rate as large as in advanced LIGO and EURO will be a 
valuable catalogue to test astronomical predictions, for example
if $\gamma$-ray bursts are associated with NS-NS or NS-BH mergers \cite{FinnCosmology}.

\subsubsection{NS-BH binaries}
These are binaries consisting of one NS and one BH and are very interesting 
from an astrophysical point of view: The initial evolution of such systems 
can be treated analytically fairly well, however, the presence of a BH
with large spin can cause the NS to be whirled around in precessing orbits 
due to the strong spin-orbit coupling. The evolution of such 
systems is really not very well understood. However, it should be possible
to use the ``point-mass" approximation in which the NS is treated as a
point-particle orbiting a BH, in getting some insight into the
dynamics of the system. The evolution will also be complicated by the
tidal disruption of the NS immediately after reaching the last stable
orbit. It should be possible
to accurately measure the onset of the merger phase and deduce the
radius of the NS to $\sim 15\%$ and thereby infer the EOS
of NS \cite{Vallisneri}.  

Advanced interferometers will be sensitive
to NS-BH binaries out to a distance of about 650 Mpc. 
The rate of coalescence of such systems 
is not known empirically as there have been no astrophysical NS-BH binary 
identifications.  However, the population synthesis models give \cite{GLPPS} a
Galactic coalescence rate in the range $3\times 10^{-7}$--$5\times 10^{-6}$~yr$^{-1}.$ 
The event rate of NS-BH binaries will be worse than BH-BH of the
same total mass by a factor of $(4\eta)^{3/2}$ since the SNR goes down
as $\sqrt{4\eta}.$ Taking these factors into account we get 
an optimistic detection rate of NS-BH of 1 to 1500 in initial and 
advanced LIGO, respectively.
 
\subsubsection{BH-BH binaries}
Black hole mergers are the most promising candidate sources for a first
direct detection of GW. These sources are the most interesting from the
view point of general relativity as they constitute a pair of interacting
Kerr spacetimes experiencing the strongest possible gravitational fields
before they merge with each other to form a single BH, and 
serve as a platform to test general relativity in the 
non-linear regime.  For instance, one can detect the scattering of 
GW by the curved geometry of the binary \cite{LBBSCQG}, \cite{LBBSPRL}, 
and measure, or place upper limits on, the mass of the graviton to 
$2.5 \times 10^{-22} $~eV and $2.5 \times 10^{-26}$~eV in ground- and 
space-based detectors, respectively \cite {WillGraviton}.
High SNR events (which could occur once every month in
advanced LIGO) can be used to test the full non-linear gravity by comparing
numerical simulations with observations and thereby gain a better understanding
of the two-body problem in general relativity. As BH binaries can be seen to
cosmological distances, a catalogue of such events compiled
by LIGO can be used to measure Cosmological parameters (Hubble constant,
expansion of the Universe, dark energy) and test
models of Cosmology \cite{FinnCosmology}.

The span of interferometers to BH-BH binaries varies from 100 Mpc 
(with the inspiral signal only) to 150 Mpc (inspiral plus 
merger signal) in initial LIGO and to a red-shift of $z=$~0.4--0.55 in 
advanced LIGO, and $z=2$ in EURO 
(cf. Fig.~\ref{fig:SourcesAndSensitivity} and \ref{fig:BinariesSNR}).  
As in the case of NS-BH binaries, here too there is no 
empirical estimate of the event rate. Population synthesis models are
highly uncertain about the Galactic rate of BH-BH coalescences and
predict \cite{GLPPS} a range of $3\times 10^{-8}$--$10^{-5}$~yr$^{-1},$
which is smaller than the predicted rate of NS-NS coalescences.
However, owing to their greater masses, BH-BH event rate in our detectors
is larger than NS-NS by a factor $M^{5/2}$ for $M \lsim 100\,M_\odot.$ The predicted
event rate is a maximum of 1~yr$^{-1}$ in initial LIGO and 500~yr$^{-1}$
to 20~day$^{-1}$ in advanced LIGO.

\subsubsection{Massive black hole binaries}
It is now believed that the centre of every galaxy hosts a BH whose mass
is in the range $10^6$--$10^9\,M_\odot$ \cite{ReesMBH}. These are termed as
{\it massive black holes} (MBH). There is now observational evidence that
when galaxies collide the MBH at their nuclei might get close enough to be
driven by gravitational radiation reaction and merge within the 
Hubble time \cite{KomossaEtAl}.
For a binary with $M=10^6M_\odot$ the frequency 
of GW at the last stable orbit is $f_{\rm LSO} = 4$~mHz,
followed by merger from 4 mHz to the QNM at 40 mHz (if the spin
of the black holes is not close to 1). This is in the frequency range of
LISA which has been designed to observe the MBH: 
their formation, merger and activity.

The SNR for MBH-MBH mergers in LISA is shown in 
Fig.~\ref{fig:BinariesSNR} These mergers will appear as the most
spectacular events requiring no templates for signal identification,
although good models would be needed to extract source parameters.
Mergers can be seen to $z\sim 30$ and, therefore, one 
could study the merger-history of galaxies throughout the Universe and
address astrophysical questions about the origin, growth and population of
MBH.  The recent discovery of a MBH binary \cite{KomossaEtAl}
and the association of X-shaped radio lobes with the 
merger of MBH \cite{MerittEtAl} has raised the optimism concerning MBH mergers and
the predicted rate for MBH mergers is the same as the rate at which
galaxies merge, about 1~yr$^{-1}$ out to a red-shift of $z=5$ \cite{Haehnelt}. 

\subsubsection{Smirches}
\begin{figure}[t]
\begin{center}
\includegraphics[width=2.5in]{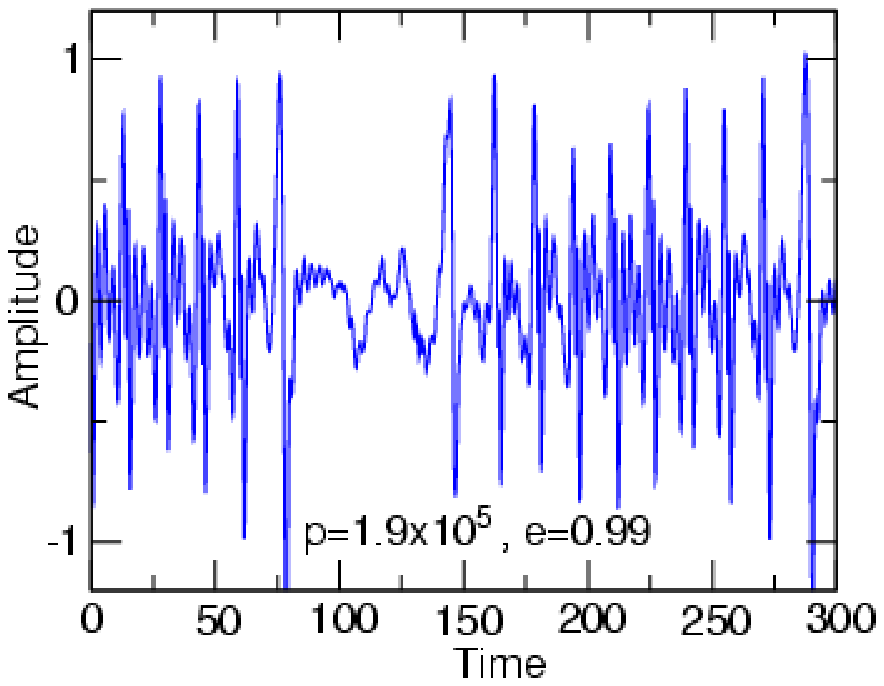}
\hskip 0.5 true cm
\includegraphics[width=2.5in]{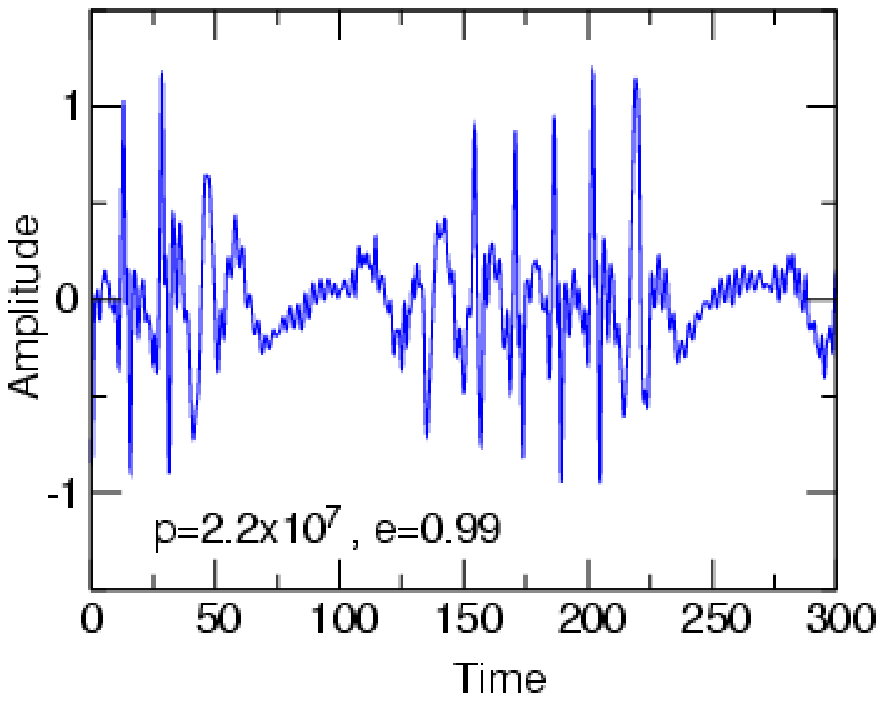}
\end{center}
\vspace{-0.5pc}
\caption{Examples of the waveform with
phase and amplitude modulation caused by spin-orbit coupling and
changing polarisation as a small BH spirals into a MBH \cite{Babak}.}
\label{fig:zoomwhirl}
\end{figure}

The MBH environment of our own galaxy is known to constitute 
a large number of compact objects and white dwarfs. Three body 
interaction will occasionally drive these
compact objects, white dwarfs and other stars into a capture orbit of the central 
MBH.  The compact object will be captured in an highly eccentric trajectory 
($e > 0.99$) with the periastron close to the last stable orbit of the MBH. 
Due to relativistic frame dragging, for each passage of
the apastron the compact object will experience several turns around the MBH 
in a near circular orbit. Therefore, long periods of low-frequency, small-amplitude
radiation will be followed by several cycles of high-frequency, large-amplitude
radiation. Waveforms from two such orbits is shown in Fig.~\ref{fig:zoomwhirl}
The apastron slowly shrinks, while the periastron remains more or less at the same
location, until the final plunge of the compact object before merger. There is a lot
of structure in the waveforms (cf. Fig.~\ref{fig:zoomwhirl}) which arises as
a result of a number of different physical effects: Contribution from higher order
multipoles, precession of the orbital plane that changes the polarisation of the
waves observed by LISA, etc.  This complicated structure smears the power
in the signal in the time-frequency plane \cite{BSS:Moriond} as compared to
a sharp chirp from a non-spinning BH binary and for this reason this {\it 
spin modulated chirp} is called a {\it smirch} \cite{BSSAndBFS}.

As the compact object tumbles down the MBH it will sample the spacetime 
geometry in which it is moving and the structure
of that geometry will be imprint in the GW emitted in the process.
By observing smirches, LISA offers a unique opportunity to directly 
map the spacetime geometry around the central object and test whether or not this 
structure is in accordance with the expectations of general relativity \cite{Ryan}. 
Indeed, according to Einstein's theory the
geometry of a rotating black hole is uniquely determined to be the Kerr metric involving
just two parameters, the mass of the MBH and its spin. Thus, the various
multipole-moments of the source are uniquely fixed once we have measured the mass and
spin of the BH.  With the observed smirch one can basically test 
(i)~whether general relativity correctly describes the spacetime region 
around a generic BH and (ii)~if the central object is indeed a BH or 
some other exotic matter.

The SNR from smirches will be between 10--50 depending on the mass of the 
central object (cf. Fig.~\ref{fig:BinariesSNR}) but it might be very difficult
to match filter them due to their complicated shapes. The
events rate is expected to be rather high. Indeed, a background population
of these smirches will cause confusion noise and only sources
in the foreground will be visible in LISA. The event
rate is as yet highly uncertain ranging from 1--10~yr$^{-1}$ within 1 Gpc \cite{Phinney}.

\subsection{Neutron stars}
Neutron stars (NS) are the most compact stars in the Universe. With a 
density of $2\times 10^{14}$~g~cm$^{-3},$ and surface gravity $\varphi
\label{sec:BHBH}
\equiv M/R \sim 0.1,$ they are among the most exotic objects whose
composition, equation-of-state and structure, are still largely unknown.
Being highly compact they are potential sources of GW.
The waves could be generated either from the various normal modes of the
star, or because the star has a tiny deformation from spherical symmetry
and is rotating about a non-axisymmetric axis, 
or because there are density inhomogeneities caused by an environment, or
else due to certain relativistic instabilities. We will consider these
in turn.
\subsubsection{Supernovae and Birth of NS}
The birth of a NS is preceded by the gravitational
collapse of a highly evolved star or the core collapse of an accreting
white dwarf. Type II supernovae (SN) are believed to result in a compact remnant.
In any case, if the collapse is non-spherical then
GW could carry away some of the binding energy and
angular momentum depending on the geometry of the collapse. It is
estimated that in a typical SN, GW might extract about $10^{-7}$ of the
total energy \cite {Muller}. The waves could come off in a burst whose frequency might
lie in the range $\sim 200$--1000 Hz. Advanced LIGO will be able to see
such events up to the Virgo supercluster with an event rate of 
about 30 per year. 

\subsubsection {Equation of State and Normal Modes of NS }
\label{sec:NSNSQNM}
In order to determine the equation of state (EOS) of a neutron
star, and hence its internal structure, it is necessary 
to independently determine its mass and radius.
Astronomical observations cannot measure the radius of a neutron
star although radio and X-ray observations do place a bound on its mass.
Therefore, it is has not been possible to infer the EOS.
Neutron stars will have their own distinct normal modes and
GW observations of these modes should resolve the matter 
here since by measuring the
frequency and damping times of the modes it would be possible
to infer both the radius and mass of NS.
The technique is not unlike helioseismology where observation
of normal modes of the Sun has facilitated insights 
into its internal structure. In other words, GW
observations of the normal modes of the NS will allow 
{\it gravitational asteroseismology} \cite{AnderssonAndKokkotas}. 

Irrespective of the nature of the collapse a number of normal
modes will be excited in a newly formed NS. The star
will dissipate the energy in these modes in the form of GWs as 
a superposition of the various normal modes and soon the 
star settles down to a quiescence state. 
Normal modes could also be excited in old NS 
because of the release of energy from star quakes.  The strongest of
these modes, the ones that are important for GW observations, are the
{\it p}- and {\it w}-modes for which the restoring forces are the
fluid pressure and space-time curvature, respectively. Both of these
modes will emit transient radiation which has a generic form of a
damped sinusoid: $h(t;\, \nu,\tau) = h_0 \exp(-t/\tau) \sin (2\pi \nu t),$ where
$h_0$ is the amplitude of the wave that depends on the external
perturbation that excites the mode and $\nu$ and $\tau$ are the
frequency and damping time of the mode, respectively, and are determined
by the mass and radius of the NS for a given EOS. 

To make an order-of-magnitude estimate let us assume that
the mass of the NS is $M_\star = 1.4\, M_\odot$ and that its radius
is $R_\star = 10\, \rm km.$
For the {\it p}-modes, which are basically fluid modes, 
the frequency of the fundamental
mode, also called the {\it f}-mode, is simply the dynamical 
frequency of the fluid, namely $\nu_f \sim \sqrt{\rho},$ where $\rho$ is
the density of the fluid. For a NS of radius $R_\star$ and mass $M_\star$
this corresponds to a frequency of 
$\sqrt{3M_\star/(4\pi R_\star^3)} \sim 3\, \rm kHz.$ If the star radiates 
all the energy deposited in the mode at a luminosity ${\cal L},$ the
damping time of the mode would be $\tau \sim E/{\cal L}.$
Since $E\propto M_\star^2/R_\star$ and 
${\cal L} \propto M_\star^2 R_\star^4 \omega^6 = M_\star^5/R_\star^5,$ we get
$\tau \sim R_\star^4/M_\star^3.$ 
Indeed, detailed mode calculations for various EOS have been 
fitted to yield the following relations for \emph{f-}modes \cite {AnderssonAndKokkotas} 
\begin{equation}
\nu_f = \left [0.78 + 1.635 \left (\frac {M_\star}{R_\star^3} \right )^{1/2}\right ] \, {\rm kHz}, \ \ \ \ 
\tau_f^{-1} = \frac{M_\star^3}{R_\star^4} \left [ 22.85 - 14.65 \frac {M_\star}{R_\star} \right ]\, \rm s,
\end{equation}
and similarly for \emph{w-}modes. 
The \emph{f-} and \emph{w}-mode frequencies nicely separate into two 
distinct groups even when considering
more than a dozen different EOS: The \emph{f}-modes are in the frequency 
range 1--4 kHz, \emph{w-}modes are in the range 8--14 kHz, and therefore, detecting
a signal at these frequencies places it in one or the other category. The
frequency and damping time, together with the relations above, can then
be used to fix the radius and mass of the star. Observing several systems
should then yield a mass-radius curve which is distinct for each EOS and
thereby helps to address the question of NS structure.

The amplitude of {\it f-} and {\it w}-modes corresponding to 12 different EOS
from NS at 10 kpc to 15 Mpc is shown in Fig.~{\ref{fig:SourcesAndSensitivity} as 
two shaded regions.
In a typical gravitational collapse the amount of energy expected to be
deposited in $f$- or $w$-modes, $\sim 10^{-8}M_\odot,$ 
makes it impossible to detect them in initial
LIGO and barely in advanced LIGO instruments, even for a Galactic source.
However, EURO should  be
able to detect these systems with a high SNR. The event rates for these
systems would be at least as large as the supernova rate, i.e. about
0.1--0.01~yr$^{-1}$ in our galaxy, increasing to 10--100 yr$^{-1}$ within
the Virgo supercluster. 

\subsubsection{Relativistic Instabilities in NS}
NS suffer {\it dynamical} and {\it secular} instabilities caused by
{\it hydrodynamical} and {\it dissipative} forces, respectively. 
What is of interest to us is the secular instability driven by
gravitational radiation. GW emission from a normal mode in a 
non-spinning NS would always lead to the decay of the mode.
However, the situation might reverse
under certain conditions: Imagine a NS spinning so fast that a 
normal mode whose angular momentum (AM) in the star's {\it rest frame} 
is {\it opposite} to its spin, appears to an {\it inertial 
observer} to be {\it co-rotating} with the spin.
In the inertial frame, GW extracts positive AM from
the mode; therefore the mode's own AM should become more 
negative. In other words, the amplitude of the mode should grow 
as a result of GW emission, and hence the instability. The energy
for the growth of the mode comes from the rotational energy
of the star, which acts like a pump field. Consequently, the star would 
spin down and eventually halt the instability. It was expected that
this instability, called the {\it CFS instability} 
\cite{Chandrasekhar}, \cite{FriedmanAndSchutz}, 
might drive the \emph{f}-modes in a NS unstable, but the star 
should spin at more than 2 kHz (the smallest {\it f}-mode frequency)
for this to happen.  Moreover, it has been shown that due to viscous 
damping in the NS fluid the instability would not grow sufficiently 
large, or sustain for long, to be observable (see e.g. Ref. \cite{AnderssonAndKokkotas}).

It was recently realized \cite{Andersson}
that modes originating in current-multipoles, as opposed to
mass-multipoles which lead to the {\it f}-mode, could
be unstable even in a non-spinning NS. These modes, called
the \emph{r}-modes, have received a lot of interest because they
could potentially explain why spin frequencies of 
NS in low-mass X-ray binaries are all clustered in a narrow range
of 300--600 Hz or why no NS with spin periods smaller than 
1.24 ms have been found.  The role of {\it r}-modes in these 
circumstances is as yet inconclusive because the problem
involves very complicated astrophysical
processes (magnetic fields, differential rotation, 
superfluidity and superconductivity), microphysics
(the precise composition of NS -- hyperons, quarks) and full non-linear
physics of general relativity. It is strongly expected
that \emph{r}-modes will be emitted by newly formed NS during the
first few months of their birth
\cite{LindblomEtAl},\cite{AnderssonAndKokkotas}. 
The frequency of these modes will be $4/3$ of the 
spin frequency of the star and might be particularly 
important if the central object in a low-mass X-ray
binary is a strange star \cite{AJK}. The radiation might 
last for about 300 years and the signal would be detectable 
in initial LIGO with a few weeks of integration.

\subsubsection{NS Environment}
A NS with an accretion disc would be spun up due to
transfer of AM from the disc. Further, accretion could 
lead to density inhomogeneities on the NS that could lead to 
the emission of GW. The resulting radiation reaction torque
could balance the accretion torque and halt the NS from spinning up.
It has been argued \cite{Bildsten}
that GW emission could be the cause for spin frequencies of NS 
in low-mass X-ray binaries to be locked up in a narrow 
frequency range of 300--600~Hz. 
It is also possible that {\it r}-modes are responsible for the locking
up of frequencies instead, in which case the waves would come off at a different
frequency \cite{AJK}. These predictions can be tested with 
advanced LIGO or EURO as Sco-X1, a nearby low-mass X-ray binary, would 
produce quite a high SNR (marked as $\star$ in Fig.~\ref{fig:SourcesAndSensitivity}).

\subsubsection{Spinning NS with Asymmetries}
Our galaxy is expected to have a population of $10^{8}$ NS  and they
normally spin at high rates (several to 500 Hz). Such a large spin must 
induce some equatorial bulge and flattening of the poles. The
presence of a  magnetic field may cause the star to spin
about an axis that is different from the symmetry axis
leading to a time-varying quadrupole moment \cite{CutlerToroidalBField}. Gravitational
waves emitted by a typical NS a distance of $r=\rm 10\,kpc$ from the Earth
will have an amplitude \cite{Thorne300Years}
$h \sim 8 \times 10^{-26} f_{\rm kHz}^2 \epsilon_{-6},$
where $f_{\rm kHz}$ is the frequency of GW in kHz
and $\epsilon_{-6}$ is the ellipticity of the star in units of $10^{-6}$. 
Fig.~\ref{fig:SourcesAndSensitivity} plots the signal strength expected 
from a NS with $\epsilon =10^{-6}$ at 10~kpc integrated over 4 months.

The ellipticity of neutron stars is not known but
one can obtain an upper limit on it by attributing the observed 
spin-down rate of pulsars as entirely due to gravitational 
radiation back reaction, namely that the change in the rotational 
energy is equal to GW luminosity. 
The ellipticity of the Crab pulsar inferred in this way is
$\epsilon \le 7 \times 10^{-4}.$ The GW amplitude corresponding
to this ellipticity is $h \le 10^{-24}.$ Noting that Crab has a spin frequency
of 25 Hz (GW frequency of 50 Hz), on integrating the signal
for $10^7$~s one obtains $h = 3.3 \times 10^{-21}\, \rm Hz^{-1/2},$ 
which is easily reachable by LIGO.  It is unlikely that the 
ellipticity is so large and hence the GW amplitude is probably much less.  
However, seeing Crab at a hundredth of this ellipticity is
quite good with advanced LIGO as indicated by a diamond in 
Fig.~\ref{fig:SourcesAndSensitivity} (Note that Crab is at 2~kpc, so with
an ellipticity of $\epsilon = 7\times 10^{-6}$ the signal strength 
would be 35 times higher than the NS line.)

\subsection{Stochastic Background}
A population of background sources \cite{GLPPS} and quantum processes
in the early Universe 
produce stochastic signals that fills the whole space.
By detecting such a stochastic signal we can gain knowledge about the
underlying populations and physical processes. A network of
antennas can be used to discover stochastic signals buried under 
the instrumental noise backgrounds. It is expected that the instrumental backgrounds
will not be common between two geographically well-separated antennas.
Thus, by cross-correlating the data from two detectors we can eliminate 
the background and filter the interesting stochastic signal. However, when detectors 
are not co-located the SNR builds only over GW wavelengths longer than twice the
distance between antennas which in the case of the two LIGO antennas means over
frequencies $\lsim 40$~Hz \cite{AllenAndRomano}.  The visibility of 
a stochastic signal integrated over a period $T$ and bandwidth $f$ only increases 
as $(fT)^{1/4}$ since cross-correlation uses a `noisy' filter. But the noise
in a bandwidth equal to frequency $f$ is $\sqrt{fS_h(f)}.$ Thus, the
signal effectively builds up as $(T/f)^{1/4}.$

\subsubsection{Astronomical Backgrounds}
There are thousands of white dwarf binaries in our galaxy with 
their period in the range from a few hours to 100 seconds. 
Each binary will emit radiation at a single frequency, but over
an observation period $T$ each frequency bin of width $\Delta f=1/T$ will
be populated by many sources. Thus, unless the source is nearby it
will not be possible to detect it amongst the {\it confusion} background
created by the underlying population.  However, a small fraction 
of this background population will be detectable as strong foreground 
sources. The parameters of many white dwarfs are known so well
that we can precisely predict their SNRs in LISA and thereby use them
as to calibrate the antenna. In the inset panel of 
Fig~\ref{fig:SourcesAndSensitivity} the curve labelled WDB is the 
expected confusion noise from Galactic white dwarfs \cite{HilsEtAl},
\cite{GLPPS}. NS and BH populations do not
produce a large enough background to be observable. Note that the white
dwarf background imposes a limitation on the sources we can observe
in the frequency region from 0.3 mHz to about 1 mHz -- the region where
we expect smirches to occur. 

\subsubsection{Primordial Background}
A cosmological background should have been created in the very early
Universe and later amplified, as a result of parametric amplification,
by its coupling to the background gravitational field 
\cite {GLPPS}.  Imprint on such a 
background are the physical conditions that existed in the early Universe
as also the nature of the physical processes that produced the background.
Observing such a background is therefore of fundamental importance as this
is the only way we can ever hope to directly witness the birth of the Universe.
The cosmic microwave background, which is our firm proof of the hot
early phase of the Universe, was strongly coupled to baryons for 300,000 years after the
big bang and therefore the signature of the early Universe is erased from it.  
The GW background, on the other hand, is expected to de-couple from 
the rest of matter $10^{-24}$~s after the big bang, and would therefore carry
uncorrupted information about the origin of the Universe.

The strength of stochastic GW background is measured in terms of the
fraction $\Omega_{\rm GW}$ of the energy density in GW as compared to the 
critical density needed to close the Universe and
the amplitude of GW is given by \cite{Thorne300Years}: $h=8 \times 10^{-19}\, \Omega_{\rm GW}^{1/2}/f,$
for $H_0=65\rm \ km\, s^{-1}\,Mpc.$ By integrating for $10^7$~s, over a bandwidth $f$, we can 
measure a background density at $\Omega_{\rm GW} \simeq 4 \times 10^{-5}$ in initial LIGO,
$5 \times 10^{-9}$ in advanced LIGO and $10^{-10}$ in LISA (cf. Fig.~\ref{fig:SourcesAndSensitivity}
dot-dashed curves marked $\Omega_{\rm GW}$).
In the standard inflationary model of the early Universe,
the energy density expected in GW is $\Omega_{\rm GW} \lsim 10^{-15},$ 
and this will not be detected by future ground-based detectors or LISA.
However, space missions currently being planned (DECIGO/BBO) to 
exploit the astrophysically quiet band of $10^{-2}$--1~Hz might detect
the primordial GW and unveil the origin of the Universe.

\section {Conventions on Source Strengths and Units}
\label{sec:conventions}
This article chiefly deals with compact objects, namely neutron stars (NS) and 
black holes (BH). Unless specified otherwise we shall assume that a NS has a 
mass of $M=1.4M_\odot$ and radius $R=10$~km, and by a stellar mass BH we shall 
mean a black hole of mass $M=10M_\odot.$ While dealing with the detectability 
of a source we shall assume that a broadband source of known phase evolution 
is integrated over a bandwidth equal to its frequency, for continuous waves
an integration time of $10^7\,\rm s,$ for stochastic signals an integration
over $10^7\, \rm s$ over a bandwidth $f,$ 
and for quasi-normal modes  an
integration over one $e$-folding time. 
These operations will convert the raw dimensionless
amplitude of GW into units Hz$^{-1/2}$, thereby allowing us to compare source
strengths with the antenna's amplitude noise spectral density $\sqrt{S_h(f)},$ which is also
measured in units of Hz$^{-1/2}.$ For a 1\% false alarm probability during the course
of observation it is typically necessary to set signal-to-noise ratio 
(SNR) threshold of about 8 in a single detector.
In a network consisting of three equally sensitive detectors,
in order that the network-SNR is 8, each instrument 
must register an SNR of $\sim 5.$ We shall therefore deem that 
a source is observable if the SNR it
produces is at least 5. The SNR achievable for point sources depends on the orientation 
of the source relative to the detector. We shall assume that sources occur with
random orientations and consider our typical source to
have an ``RMS'' orientation. The amplitude of a source with an ``RMS'' orientation
is smaller than an optimally oriented source by a factor of $5/2$ \cite{DIS2}.

We shall assume a flat Universe with a cold dark matter density of
$\Omega_M = 0.3,$ dark energy of $\Omega_\Lambda = 0.7,$ and a
Hubble constant of $H_0 = 65\,{\rm km\, s^{-1}\,Mpc^{-1}}.$
We shall use a system of units in which 
$c=G=1,$ which means $1\,M_\odot\simeq 5 \times 10^{-6}\,{\rm s} \simeq 1.5\, {\rm km},$
$1\,{\rm Mpc} = 10^{14}\, {\rm s}.$ 

There have been many reviews and books 
\cite{CutlerAndThorne}\cite{GLPPS}\cite{SchutzCQG}\cite{IyerAndBhawal}\cite{SchutzGRBook}
\cite{LasotaAndMarck}
on this topic to which we refer the reader for further reading. 
Our depiction of the noise spectral density and source strengths in 
Fig.~\ref{fig:SourcesAndSensitivity} is motivated by Ref.~\cite{CutlerAndThorne}.

\section*{Acknowledgments} For useful discussions I would like to thank 
Nils Andersson, Kostas Kokkotas, Leonid Grishchuk and Bernard Schutz.

\vspace{\baselineskip}
{}

\endofpaper
\end{document}